\documentclass{skaox2006}
\usepackage{graphicx}
\usepackage{natbib}

\def\HIm{\ifmmode \hbox{\scriptsize H\kern0.5pt{\footnotesize\sc i}}\else H\kern1pt{\small I} \fi}
\def\HIIm{\ifmmode \hbox{\scriptsize H\kern0.5pt{\footnotesize\sc ii}}\else H\kern1pt{\small II} \fi}

\begin{document}
  \title{Working Group Summary - Local Universe}

  \author{E. Tolstoy\inst{1}
         \and
          G. Battaglia\inst{2}
         \and
         R. Beck\inst{3}
         \and
         A. Brunthaler\inst{3}
         \and
         A. Calamida\inst{2}
         \and
         G. Fiorentino\inst{1}
         \and \\
         J.M. van der Hulst\inst{1}
         }

  \institute{    Kapteyn Astronomical Institute, University of Groningen, Netherlands
    \and
    European Southern Observatory, Garching, Germany
    \and
    Max Planck Institute for Radio Astronomy, Bonn, Germany
            }
  \abstract{
This is a summary of the discussions that took place in the working
group dedicated to studies of the Local Universe. The authors are
listed in alphabetical order, after the working group organiser, 
and are those who gave a presentation
during the week in Crete; their contributions are incorporated here.
During the group discussions we considered the various synergies that
exist between future studies of individual stars and star formation
regions at optical/IR wavelengths that will be possible with the E-ELT
and those of the molecular and neutral gas in similar regions that
will be possible with ALMA and SKA.  The primary emphasise was on star
formation; both on large and small scales. New facilities will allow
more detailed insights into the properties of our own Galaxy and also
allow us to make detailed comparisons with a range of more distant
systems all forming stars at different rates from different
initial conditions (e.g., metallicity) and with different spatial 
distributions.
}
  \maketitle
%
%
\section{Introduction}

The E-ELT will offer outstanding possibilities for optical/IR imaging
and spectroscopy over small fields of view (typically
$\le$50~arcsecs) with uniquely high spatial resolution at these
wavelengths (few mas). This is ideal for detailed studies of
individual stars and ionised gas in our Galaxy and also in other
galaxies out to and beyond the distance of the Virgo cluster. What is
currently lacking in this field is detailed studies of a range of
galaxy types in different environments (e.g., field, group, cluster). 
Detailed star formation histories and
spectroscopic studies of individual stars are presently restricted to
the Local Group; in some cases to within the halo of the Milky Way.

The Square Kilometre Array (SKA) will offer a unique new ability to
map at radio wavelengths with great sensitivity over 
large
field of views ($\sim$1~degree at 1~Ghz, and much larger at
lower frequencies) with a relatively high spatial resolution
(0.2$-$20~arcsec). This is ideal for studies of neutral gas, thermal and and
non-thermal emission in our Galaxy, the Local Group and well
beyond. One crucial aspect of SKA 
is that the break through will not come {\it only} from
looking at systems further away but also allowing more detailed
observations of the nearby Universe. Observations will be possible
with higher spatial resolution to look
at small structures and higher surface brightness 
sensitivity to observe in more detail faint 
diffuse structures that are believed to surround galaxies. 
The enhanced sensitivity of a SKA even operating
at a 10\% of its full capability will lead to significantly more
sensitive maps of the gas content for a range of different galaxy
types in the nearby universe (and well beyond).

  \begin{figure}
  \centering
\resizebox{\hsize}{!}{\includegraphics{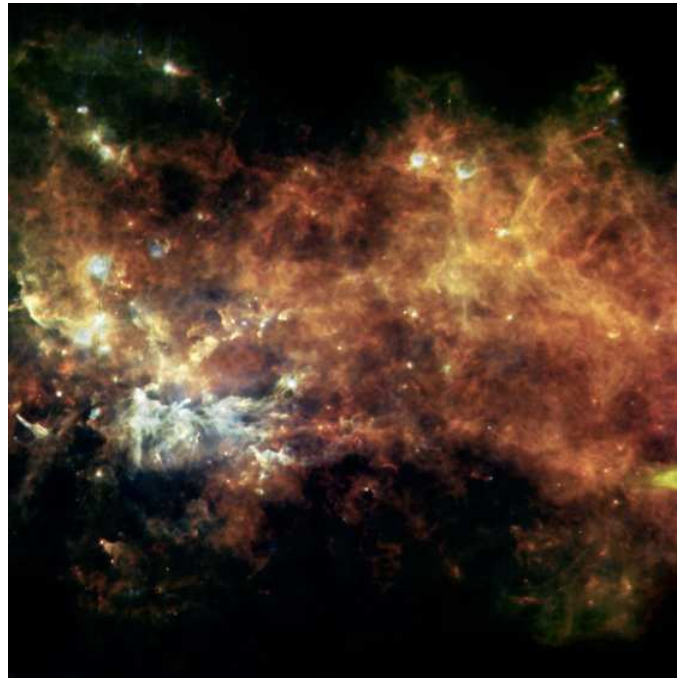}}
     \caption{
European Space Agency's Herschel infrared space observatory infrared image of
the Galactic Vulpecula star forming region. It shows us how stars are formed inside filaments of glowing gas and dust.
credit: ESA/Hi-GAL Consortium.}
\label{herschel}
  \end{figure}

Equally exciting are the detailed studies of star formation processes, but
we did not consider these closely as
we lacked someone with the necessary experience in the group. It is clear that
satellites such as Herschel, 
the James Webb Space Telescope (JWST) and also 
the future ground based sub-millimeter array ALMA will offer
extraordinary sensitivity to allow exceptionally detailed studies of
the properties of molecular gas (see Figure~\ref{herschel}) 
which will allow us to almost watch
stars form in the nearby Universe on a spatial and spectral resolution
scale similar to that of the E-ELT. This will
undoubtedly lead to samples of objects that will require spatial and spectral 
high resolution follow-up with ELT instrumentation and maybe also SKA.

At first glance it maybe hard to envisage obvious synergies between SKA \& E-ELT as the spatial
scales observed are very different. 
However, both wavelength regions provide unique and different information and bringing them together 
makes a powerful combination for tackling the complex processes of star formation.  Gas is the raw
material for star formation; looking at \HIm shows where the stars may
form, and for how long; molecular gas facilitates the study of the
actual process of gas turning into stars; and optical/IR imaging and
spectroscopy allow you to study all the stars that have ever been
formed in a stellar system (one by one if the system is close enough).
This makes the detailed study of star formation processes 
an extremely interesting synergy between SKA and E-ELT (and also other
major observatories). The careful and sensitive study of magnetic
fields is another field that will benefit from the synergy between
radio and optical wavelengths, especially if polarisation is considered.
Studying individual objects such as masers and 
variable stars also allows highly accurate mapping of galactic
structure and also an accurate distance scale.

The combination of SKA and E-ELT will thus allow us to understand the
effect that stars have on their Interstellar Medium (ISM) and also the
effect that ISM has on stars. It is a circular process and many
aspects are currently poorly understood and yet assumptions and
parameterisations of this process permeate all aspects of the
interpretation of present day large scale galaxy surveys out to the
highest redshifts and also all models of the formation and evolution
of galaxies.

\section{A Description of general  science areas}

During the group discussions we considered the various synergies that
exist between studies of individual stars and star formation regions
at optical/IR wavelengths that will be possible with the E-ELT and
those of the molecular and neutral gas in similar regions that will be
possible with SKA, and to a lesser extent ALMA and JWST.  The primary
emphasise was on star formation; both on large and small scales. New
facilities will allow more detailed insights into the properties of our
own Galaxy and also allow us to make detailed comparisons with a range
of more distant systems all forming stars at different rates and with
different distributions.

\subsection{Resolved Stars}

The combination of optical spectroscopy and imaging of individual
stars in the nearby Universe can teach us in detail about the dominant
processes in galaxy formation and evolution through-out the history of
the Universe. This is one of the key science areas identified in the
E-ELT science case (and also in all other ELT projects). Studies which
resolve individual stars will always benefit dramatically from large
increases in spatial resolution and flux sensitivity.

  \begin{figure}
  \centering
\resizebox{\hsize}{!}{\includegraphics{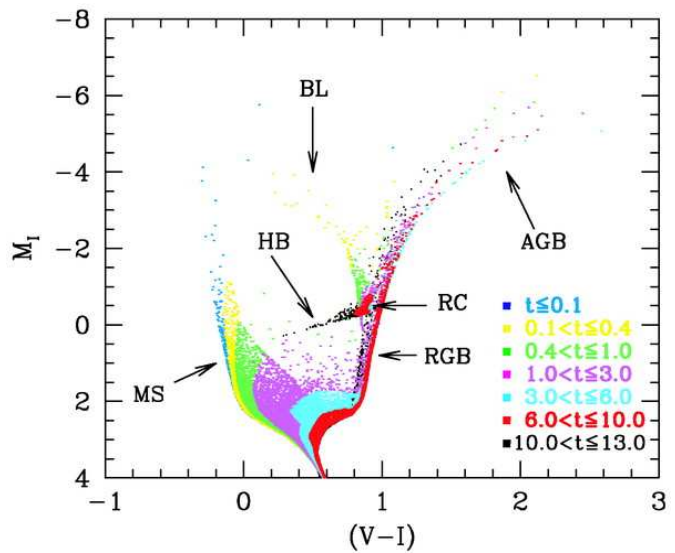}}
     \caption{
From \citep{Aparicio04} Synthetic CMD computed using constant SFR
from 13 Gyr ago to date and metallicity linearly increasing. Stars in
different age intervals are plotted in different colours. The colour
code is given in the figure, in gigayears. Labels indicate the
evolutionary phase.
}
        \label{cmd-fig}
  \end{figure}

\subsubsection{Star Formation Histories}

Within the Local Universe galaxies can be studied in great detail star
by star.  At the beginning of the past century stars were found to
group themselves in temperature-luminosity space (observed as colour
and magnitude) called the Hertzsprung-Russell Diagram; and it was
later understood that the positions reveal the evolutionary sequences
of stars of different ages and masses.  It is now well established
that the resolved Colour-Magnitude Diagrams (CMDs) of stellar systems
retain detailed information about the past star formation history
(SFH). They preserve the imprint of fundamental evolutionary
parameters such as age, metallicity, and initial mass function in such
a way that it is often possible to disentangle them (see
Figure~\ref{cmd-fig}).  Because low mass stars have such long lifetimes,
for stars with mass $<0.8~M_\odot$ this can be longer than the present
age of the Universe. Using their photospheres we can probe the ISM
from which they formed, and thus measure how the chemical enrichment
of the gas in a galaxy changes throughout its history of star
formation. This provides detailed information about how enrichment
processes such as supernovae explosions and stellar winds affect the
evolution of the galaxy and its environment from the early Universe
until the present day.  The CMD synthesis method is well established
as the most accurate way to determine SFHs of galaxies back to the
earliest times \citep[e.g.,][]{Aparicio04, Tolstoy09}.  It clearly
depends on the ability to take deep sensitive images with high spatial
resolution and to be able to accurately photometer large numbers of
stars with a large range of luminosity and colour down to the oldest
Main Sequence Turnoffs.

Star formation in our Galaxy occurs over a restricted metallicity
range. If we wish to study the effects of much lower metallicities,
which more closely match the physical conditions of the early universe
we have to look beyond the Milky Way. To see the most active star
forming systems (starburst, or Blue Compact Dwarfs, BCDs) 
then we have to look well beyond
the Local Group (e.g., NGC~1569 or M~82).

\begin{figure}
  \centering
\resizebox{\hsize}{!}{\includegraphics{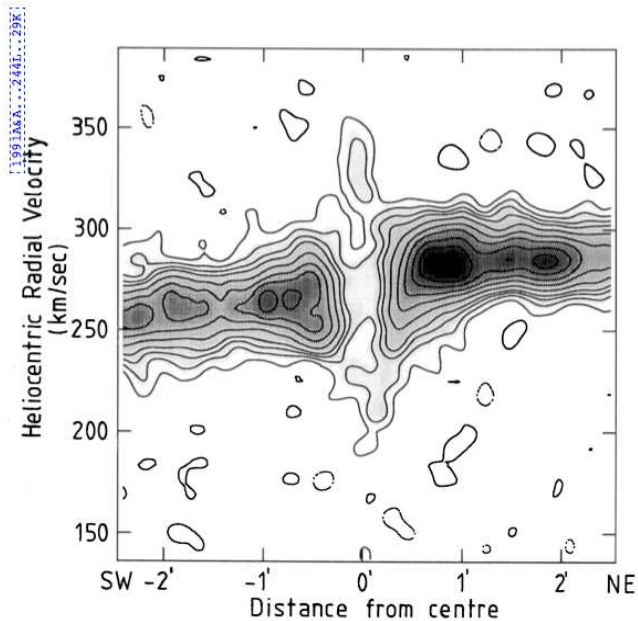}}
     \caption{
From \citep{Kamphuis91}, position-velocity map centred on an \HIm super bubble 
associated with the \HIIm region NGC~5462 in M~101. Note the clear velocity 
symmetry of the expanding shell centered on the large \HIm hole in NGC~5462.
Such observations probe both the amount of ejected \HIm and associated kinetic
energy budget
}
        \label{m101}
  \end{figure}

These kinds of studies can be carried out for nearby resolved galaxies
in the entirety to study both recent star formation and also the
entire star formation history going back to the earliest times. 
It is possible to 
look at the properties of gas and star formation in actively star
forming systems and compare this in detail to more quiescent systems.
The features in \HIm gas properties (e.g., holes, detailed
kinematics, see also \citealp{Boomsma08}) 
can be linked in more detail to the stellar properties -
if individual stars can be detected and accurately photometered. It
will be possible to verify if the local \HIm distribution and
kinematics are consistent with one or more Supernovae explosion, or
stellar winds (see, Figure~\ref{m101})
or if an external explanation always has to be found.

We may hope to distinguish the fundamental differences between
Elliptical galaxies and galaxies with disks, and variations of disk
type and the importance of bulges. That is to derive a physical
understanding of the Hubble sequence. The most serious lack at present in
furthering this kind of science case is detailed observations and
measurements of resolved stellar populations (e.g., CMD analysis)
of galaxies more distant
than the Local Group. The 
ELT project has resulted
in a number of detailed simulations of what to expect, so far SKA has
not done this (for Local Universe studies), but see Table 1, which gives
an overview of SKA capabilities in the Local Universe for a range of 
different baselines.

  \begin{figure}
  \centering
\resizebox{\hsize}{!}{\includegraphics{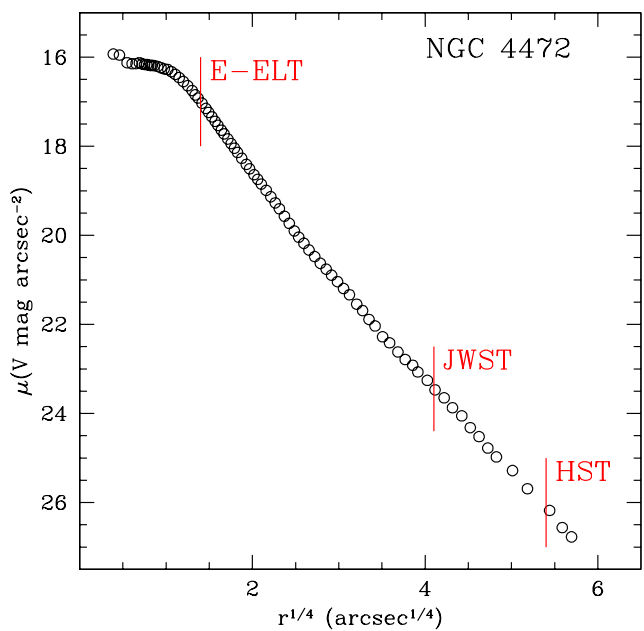}}
     \caption{From G. Fiorentino, 
the surface brightness limit at which stars can be
resolved and photometered for E-ELT, JWST and HST in 
a bright Virgo Elliptical galaxy (NGC~4472).
}
        \label{Giul-fig}
  \end{figure}

For E-ELT imaging the point spread function is likely to be highly complex
and detailed simulations have been carried out as part of the
E-ELT/MICADO Phase A study as to how these images can be analysed (by
A. Deep and G. Fiorentino). These facilities
will also only operate at red wavelengths which also has an effect on 
the accuracy of the SFH
analysis. It is predicted from simulations (Deep et al., in prep), 
that E-ELT CMDs of a
reasonable accuracy can be achieved at the distance of Virgo for a
SB(V)=19 mag/arcsec$^2$. This means a distance of 
$\sim$15~arcsec from the centre
of the giant elliptical galaxies, close to the core radius. 
Making it possible for the first
time to look in detail at the resolved stellar population of a giant
Elliptical near the central (high surface brightness) 
regions (see Figure~\ref{Giul-fig}).  
Test observations have also been carried out using MAD
to understand these effects on the sky (Fiorentino et al., in prep.).


\subsubsection{Kinematics \& Metallicity}

It has long been know that combining kinematics and metallicities is
the only means we have to separate the diverse stellar populations in
the Solar Neighbourhood 
\citep[e.g., as first determined by][]{ELS62}.
Stars can be split up into disk and halo components on the basis of
their kinematics and these subsets can then be studied independently.
Kinematics and metallicity indicators (lower resolution spectroscopy,
R$=5000-10~000$) have been more recently updated and shown to be
excellent tools for disentangling the properties of complex stellar
systems like our own Galaxy \citep[e.g.,][]{Venn04} and also other
nearby galaxies with sufficiently large samples of stellar spectra have
been found to be multi-component stellar systems
\citep[e.g.,][]{Battaglia07}.  These studies have shown how the
properties of stellar populations can vary spatially and
temporally. This leads to important constraints to theories of galaxy
formation and evolution.

Dwarf spheroidal galaxies are currently the closest early-type
galaxies that contain sufficiently large numbers of well-distributed RGB stars
to provide useful kinematic and metallicity probes. Moreover, dSphs
are considered to be ideal environments to search for dark matter
because of their very large dynamical mass-to-light ratios that cannot
be explained by the little luminous matter dSphs contain, assuming
that these galaxies are in equilibrium.  The velocity dispersion of
individual stars can be used to determine the mass of the galaxy. It
is also possible to determine metallicities for the same stars. This
allows a more careful distinction of the global properties based on
structural, kinematic, and metallicity information 
\citep[e.g.,][]{Battaglia07, Battaglia08}.

\begin{figure}
  \centering
\resizebox{\hsize}{!}{\includegraphics{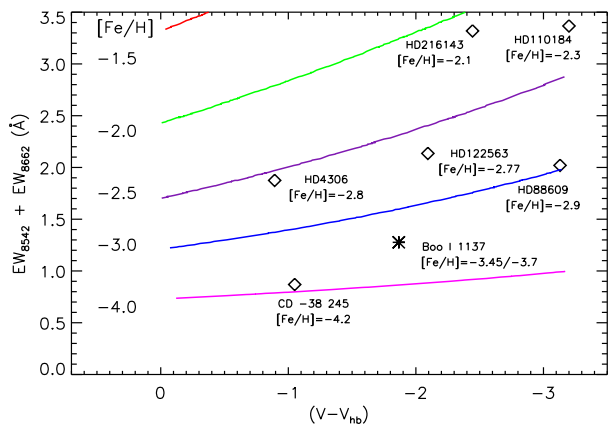}}
     \caption{From Starkenburg et al. (2010), 
new CaT calibration (coloured lines) and a sample of well-studied RGB stars in the Milky Way halo (black diamonds) and one extremely low-metallicity star in Boötes I (black asterisk).
}
        \label{stark-fig}
  \end{figure}

Another aspect of the surveys of individual RGB stars in dSphs has
been the determination of metallicity distribution functions (MDFs),
typically using the CaII triplet metallicity indicator 
\citep[e.g., most
recently][]{Stark10}.  This uses the
empirical relation between the equivalent width of the CaII triplet
lines and [Fe/H].  The accuracy of this relation, calibrated
on stellar clusters, has been shown to fail at low metallicities, [Fe/H]$<
−2.5$, and this deviation has been quantified. There is now a new 
calibration extended down to [Fe/H]$\sim -4$, making the Ca II
indicator the most useful and efficient metallicity indicator over the entire
range of stellar metallicities (see Figure~\ref{stark-fig}).

There is no reason that these techniques cannot be applied with E-ELT
to more distant galaxies.  Simulations of spectroscopic observations
with E-ELT have been made (by G. Battaglia) to look for detailed
metallicity distributions in galaxies at a range of different
distances and also to study their kinematics. 
These simulations, assuming an IFU spectrograph, 
suggest that detailed studies of a large sample of individual stars
can be carried out to $\sim$4~Mpc (e.g., Cen~A) in reasonable
observing times.  The shape of the PSF appears to be a limiting factor
in the results of the simulations, because in the CaT region it is
currently predicted to have an extremely low strehl.

These kinds of detailed kinematic studies are just starting to be
carried out in dwarf galaxies with gas, but these will be limited by
sensitivity. This means that E-ELT will greatly increase the sample,  
and range of morphological types, for
which this is possible. SKA
will greatly increase the sensitivity of existing \HIm maps and create new
ones for more distant systems, including Elliptical galaxies many of
which appear to possess \HIm \citep[e.g.,][]{Oosterloo07, Serra07}. 
SKA will in addition locate new systems as
surveys may reveal small, low \HIm mass galaxies where star formation
is progressing very slowly. Local Group Dwarfs will be imaged in \HIm
with unprecedented detail and surpass the interesting results in, for
example Leo T \citep{2008MNRAS.384..535R}.

\subsubsection{Stellar Abundances}

Detailed Stellar abundances of a range of different chemical elements in
individual stars requires 
high resolution spectroscopy, (R$>20~0000$).
Spectroscopic studies using large
ground-based telescopes such as VLT, Magellan, Keck, and HET have
allowed the determination of abundances and kinematics for signiﬁcant
samples of stars mostly in the Milky Way and the most nearby dwarf
galaxies \citep[e.g.,][]{Tolstoy09}. 

The detailed chemical abundance patterns in individual stars of a
stellar population provide a fossil record of chemical enrichment over
different timescales.  As generations of stars form and evolve, stars
of various masses contribute different elements to the system, on
timescales directly linked to their mass. Of course, the information
encoded in these abundance patterns is always integrated over the
lifetime of the system at the time the stars being studied were born.
Taking high resolution spectra of individual stars allows us to use a
variety of chemical elements as detailed indicators of a variety of
physical processes of chemical enrichment (supernovae; AGB stars;
planetary nebula; galactic winds etc) and how they varied over time
\citep[e.g.,][]{McWilliam97}.  Using a range of stars as tracers provides
snapshots of the chemical enrichment stage of the gas in the system
throughout the SFH of the galaxy. 
These studies require precise measurements of elemental
abundances in individual stars, and this can only be done with
high-resolution and reasonably high signal-to-noise spectra.  It is
only very recently that this has become possible beyond our Galaxy. It
is efﬁcient high resolution spectrographs on 8$-$10m telescopes that
have made it possible to obtain high-resolution (R $>$ 40,000) spectra
of RGB stars in nearby dSphs and O, B, and A super-giants in more
distant dIs.  Looking exclusively at young objects, however, makes it
virtually impossible to uniquely disentangle how this enrichment built
up over time.

E-ELT can push these possibilities for abundance surveys
to the edges of the Local Group and hopefully beyond.  The
study of resolved stellar populations within the Local Group and out
to distances well beyond is one of the key science cases for the ESO
E-ELT.  As such it has been simulated both for imaging
and spectroscopy as part of the ESO Science Working Group and also as
part of many of the E-ELT phase A instrument studies 
(e.g., MICADO, imaging; HARMONI, spectroscopy).

  \begin{figure*}
  \centering
\resizebox{\hsize}{!}{\includegraphics{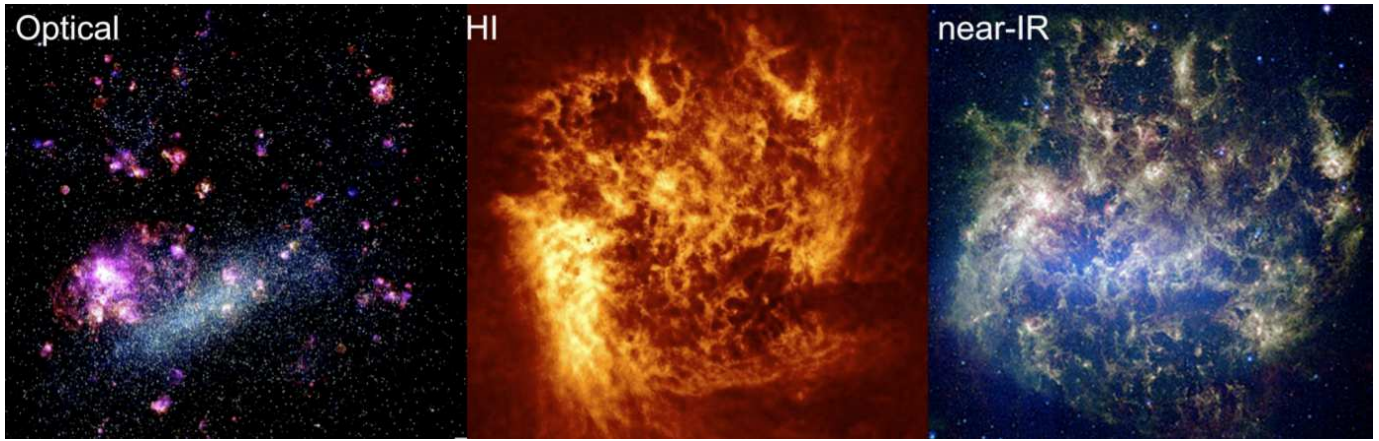}}
     \caption{
Images of the Large Magellanic Cloud in Optical, Radio \& Infra-red wavelengths,
illustrating the amount of detail SKA will provide for galaxies in the Local 
Universe.
}
        \label{lmc}
  \end{figure*}

\subsection{Gas}

\begin{figure}
  \centering
\resizebox{\hsize}{!}{\includegraphics{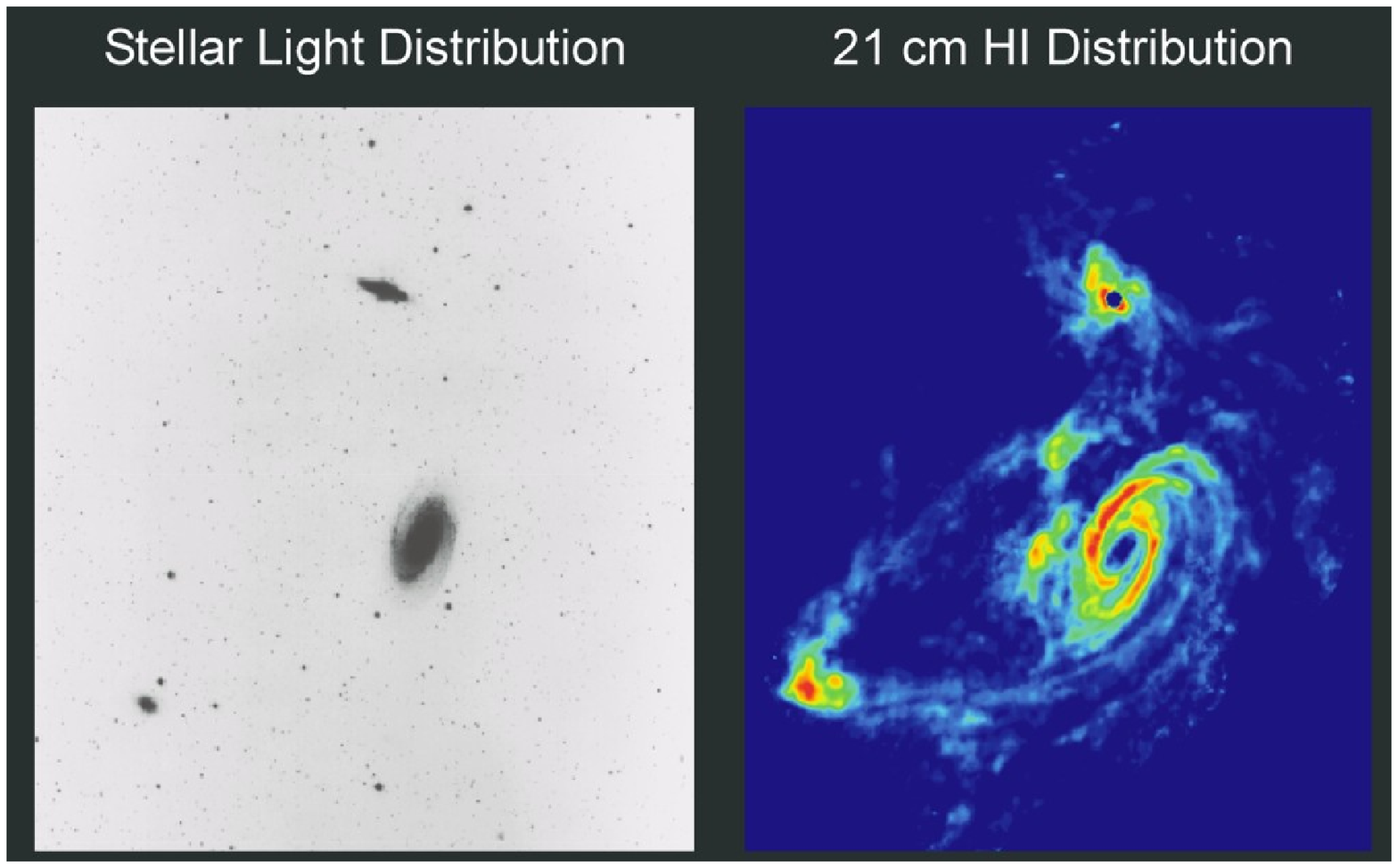}}
     \caption{
Optical and \HIm image of the M~81 group, illustrating the striking
difference between the distributions of light and \HIm
\citep[][]{1994Natur.372..530Y}
}
        \label{M81-fig}
  \end{figure}

Stars by themselves are excellent probes of global processes going
back to the formation epoch of any stellar system.  These kinds of
studies however do not tell you much about {\it how} or {\it why} the
stars formed. Looking at the stars or the gas properties of a galaxy
can tell a very different story 
(e.g., M~81 and its environs \citep{1994Natur.372..530Y}, see also 
Figure~\ref{M81-fig}).
This is because stars have
typically been forming over the age of the galaxy and the gas may be
disrupted at any point and so change the evolutionary path of a
galaxy. How the gas turns into stars
may be deduced from the properties of young stellar populations. This
gets more indirect the older the stellar population, so this remains 
a study that can only be done at present times. 
The role of feedback in galaxy formation 
appears increasingly important for coupling theory and observations
\citep[e.g.,][]{2010MNRAS.406.2325O, 2009ApJ.707..954P} 
while the evidence for coupling between 
gas inflows and/or outflows and star formation activity in galaxies
is slowly emerging from current observations of \HIm in and around galaxies
\citep[e.g.,][]{Sancisi08}.

 \begin{table}
    \caption[]{SKA capabilities for \HIm imaging in the Local Universe 
     \citep[van der Hulst, priv.comm., see also][]{vdh04}}
       \label{tab:thijs}
   $$
       \begin{array}{cllllll}
          \hline
          \noalign{\smallskip}
           Baselines & Dist. & noise & resol. & resol.&   M(\HIm) & N(\HIm) \\
          \noalign{\smallskip}
            \le  km        & Mpc   & \mu Jy  &  parsec   & arcsec    & 5   
\sigma  & 1 \sigma \\
\\
          \hline\\
          \noalign{\smallskip}
2 & 1  & 2.5 & 108.0  &18.0   & 2.5E+02 & 4.0E+17 \\
   & 7  & 2.5 & 648.9  &18.0   & 9.2E+03 & 4.0E+17 \\
   & 13 & 2.5 & 1182.3 & 18.1  & 3.1E+04 & 4.0E+17 \\
   & 19 & 2.5 & 1717.3 & 18.1  & 6.5E+04 & 4.0E+17 \\
6  &1   &2.5  & 36.0   &6.0    & 2.1E+02 & 3.6E+18 \\
    &7   &2.5 & 216.3  &6.0    & 7.6E+03 & 3.6E+18 \\
    &13  &2.5 & 394.1  &6.0    & 2.6E+04 & 3.6E+18 \\
    &19  &2.5 & 572.4  &6.0    & 5.4E+04 & 3.6E+18 \\
50  &1   &1.5 & 4.3    &0.7    & 1.7E+02 & 1.5E+20 \\
    &7   &1.5 & 26.0   &0.7    & 6.1E+03 & 1.5E+20 \\
  &13 &1.5 &47.3& 0.7 &2.1E+04 &1.5E+20 \\
  &19 &1.5 &68.7& 0.7 &4.3E+04 &1.5E+20 \\
          \noalign{\smallskip}
\\
          \hline
       \end{array}
   $$
\begin{list}{}{}
\item[$^{\mathrm{a}}$] assumed $\Delta$$V$ = 50 km/s for $M(\HIm)$ and $N(\HIm)$.
\item[$^{\mathrm{b}}$] $M(\HIm)$ and $N(\HIm)$) are measured per resolution element
assuming a 12 hour integration.
\end{list}
 \end{table}

Studying the neutral hydrogen (\HIm) gas is relevant in two ways: (i)
high spatial resolution imaging will provide a detailed picture of the
distribution and kinematics of the high column density \HIm on scales
of a few tenths of parsecs, i.e.  the scale of OB associations and
star clusters \citep[e.g., see][]{2004NewAR..48.1305B}; and 
(ii) low spatial resolution imaging will reveal the large scale distribution
and kinematics of extended, low column density \HIm. Table 1 provides
an overview of the limiting \HIm masses and column densities for
different distances in the nearby universe. Note that there is a range
of spatial resolutions that will be available with SKA, depending on what 
fraction of the collecting area 
is being used. This illustrates the flexibility of the instrument
and the fact that useful observations of the nearby Universe can start
very early in the construction phase.  More detailed information about
the capabilities of the SKA for \HIm imaging can be found in 
van der Hulst et al. (2004).

For comparison,
current synthesis instruments provide at best column density sensitivities 
of a few $10^{20}$ cm$^{-2}$ with angular resolutions of $\sim 6''$ 
\citep[e.g.,][]{2008AJ.136..2563}. Though such observations
begin to unravel relations between the gas and star formation activity
\citep[e.g.,][]{2008AJ.136.2846}, the linear resolution (a few hundred 
parsecs) and sensitivity ($\sim 10^{21}$ cm$^{-2}$) are far from adequate.
On these spatial scales additional information about the molecular gas, traced 
with the CO lines, which can be observed by ALMA, will provide 
complementary information.
At present the LMC/SMC is one of the few systems which are near enough to carry
out detailed observations in \HIm, molecular gas, radio continuum,
ionised gas and individual stars (see Figure~\ref{lmc}). However, it is very 
hard to generalise about
this system as it is clearly undergoing a very strong interaction with
the Milky Way. This is the kind of study that the combination of SKA
and E-ELT will allow us to carry out for a range of different galaxies
large and small in nearby groups (e.g., Sculptor group) 
and also in large clusters such as Virgo and Coma where we
start to find large Elliptical galaxies in \HIm.
We will also be able to carry out much more
detailed surveys of galaxies in the Local Group which are currently
not accessible to detailed high resolution spectroscopy or very deep
imaging (e.g., M31/M33 and also a range of large and small dwarf galaxies).

The local Universe, going out to Virgo cluster (at $\sim$20~Mpc) will
provide uniquely detailed insight into the formation and evolution of
both gas and the relation to stars for a large range of galaxy types.
Further afield even the SKA resolution will not be adequate to examine 
details on the tens of parsec scales. Only in the local Universe will it be 
possible to connect 
global \HIm signatures: interaction, global outflow, accretion,
presence of \HIm companions 
\citep[e.g. as demonstrated in][]{Sancisi08} 
with the detailed gas properties on
small scales. SKA will thus provide crucial information. 
Table 1 demonstrates that such \HIm signatures can easily be recognised
in galaxies well beyond the Virgo cluster.

\subsection{Magnetic Fields}

Studying synchrotron emission in the Milky Way and beyond will
provide insights into galactic-scale magnetic fields and the effects
of small-scale magnetic fields on ISM processes and star formation.
Polarised radio emission is a tracer of ordered fields in the warm
ionised ISM, fields in gas clouds can be studied by Zeeman
measurements of radio lines, and IR polarimetry gives access to
magnetic fields in dust clouds. Polarimetry for the E-ELT is crucial
to exploit these synergies with SKA.

The tight correlation between infrared and radio continuum
intensities tell us that magnetic fields and star formation are
closely related, but we are far from understanding how this coupling
works. Are magnetic fields amplified by turbulent gas motions
excited by supernova remnants, by stellar outflows or by shearing
and compressing motions due to the gravitational distortions around
spiral arms and bars? MHD modelling shows that magnetic fields are
amplified to the equipartition level with turbulent kinetic energy
within short time and then affect the gas flows 
\citep{avillez05}. The structure of collapsing
molecular clouds and the formation of star clusters can also be
significantly modified by magnetic fields 
\citep{price08}. Observational tests need high spatial
resolution and 
sensitivity in selected star-forming regions of the Milky Way. The
SKA will provide the radio polarization and Zeeman data, ALMA the
molecular clouds, while the E-ELT can observe the dust polarisation.
The shear and compression of the gas flows as measured from the
kinematics of ionised gas (E-ELT), HI (SKA) and CO (ALMA) can be
compared with the magnetic field strength (SKA). This will allow us
to directly observe the interaction between gas and magnetic fields.

One of the open questions is whether ambipolar diffusion or turbulence
governs the field evolution in collapsing molecular clouds 
\citep{crutcher09}. High-resolution IR and sub-mm polarimetry
and Zeeman radio data are needed here. Radio continuum can provide
information about the ambient magnetic field and the amount of magnetic
braking of the rotating cloud. Is also possible that rotating clouds
drive dynamos which would further enhance the field.

Combined SKA and E-ELT observations are also of great interest in
low-luminosity dwarf galaxies, as described in section 2.1.2, and especially
in BCD and dwarf irregular galaxies where active star formation is ongoing. 
Is there a threshold in gas density
for the onset of star formation? How strong are environmental
effects? Are outflows more frequent in galaxies with lower
gravitational potential? The answers to these questions need
combined efforts from the radio and optical ranges. Magnetic fields
mapped by radio polarisation may show us interactions with the
environment and outflows. On the other hand, the threshold for field
amplification could well be higher than that for star formation, so
that only thermal radio emission is observable. Galaxies without
magnetic fields probably have a steeper IMF and can serve as a
template for the first star-formation regions in the early Universe.

The generation of large-scale magnetic fields by the mean-field
dynamo needs differential rotation and turbulence, both of which are
weak in low-luminosity dwarf galaxies. Kinematic information from
a sample of dwarfs with different star-formation rates obtained
with the E-ELT can provide important parameters to understand how the
dynamo works.

Other classes of galaxy where little is known about their magnetic
fields are giant Ellipticals and Dwarf Spheroidals (dSph). The 
lack of any ongoing star formation 
makes them especially
ideal candidates to search for
synchrotron emission from secondary particles produced by decaying
dark matter \citep{cola07}, but only if magnetic
fields exist. Field amplification needs turbulence which can be provided
by turbulent wakes in the gas or by the magneto-rotational instability.
The detection of warm or hot gas and kinematic information, which needs
the E-ELT sensitivity and resolution, are crucial here.

In Elliptical galaxies with hot gas, relics of a large-scale magnetic field from
the period of strong star formation may have survived. A sample of galaxies
with different star-formation histories, as measured from their stellar
populations, will allow us to measure the lifetime of magnetic fields
after star formation ended. 

\subsection{Astrometry}

Many astrophysical properties of objects in star forming regions, like
the size, luminosity, and mass, depend strongly on the distance. Therefore,
a true understanding of star formation and a detailed comparison of 
theoretical models and observations is only possible if accurate distances
are known to star forming regions. Current galactic distance estimates are 
often affected by dust obscuration (photometric distances) or highly model
dependent (kinematic distances). A completely unbiased method to estimates 
distances is the trigonometric parallax, which requires extremely precise
astrometric measurements in the range of a few micro-arcseconds ($\mu$as). 
While {\it Gaia} will measure parallaxes of a billion stars in the Galaxy, 
it will not probe parts of the Galaxy that are obscured by dust, i.e. large 
parts of the plane of the Milky Way beyond 1 or 2 kpc inward from the Sun 
and in particular not the deeply obscured regions where new stars are forming.
However, radio observations at cm-wavelengths are not hindered by dust and can 
provide a view of the Milky Way that is complementary to {\it Gaia}.

Currently, astrometric radio observations of methanol and water masers in star
forming regions with Very Long Baseline Interferometry (VLBI) can reach 
parallax accuracies of up to 6\,$\mu$as 
\citep{Reid09a, Hach09}. 
Therefore, these observations have 
the potential to measure accurate distances to most Galactic star forming 
regions, to map the spiral structure of a large part of the Milky Way,
and to determine important parameters such as the rotation velocity and the
distance to the galactic center with high accuracy \citep{Reid09b}.

SKA will have even better astrometric capabilities than current VLBI 
arrays. Since systematic errors from the atmosphere or the array 
geometry (e.g. antenna positions, earth orientation parameters) scale directly 
with the angular separation of the reference and target sources on the sky, 
it is essential to use 
reference sources that are as close as possible. The superior 
sensitivity will allow the use of much weaker (and therefore much closer) 
background reference sources.  With parallax accuracies approaching $\sim$ 
1\,$\mu$as, the SKA can measure distances to sources out to 10 kpc with 1\% 
accuracy.

SKA in phase~I can already 
observe the important 6.7 GHz methanol maser line. This
line is a tracer of high mass star formation, widespread in the Galaxy, and 
has been already used for astrometric observations 
\citep{Rygl10}.
In phase~III, SKA will also cover the 22 GHz water maser line, the 
strongest maser line in star forming regions. Furthermore, the large continuum 
sensitivity of the SKA allows the observation of radio stars at much larger 
distances than the few hundred parsec currently possible 
\citep{Menten07, Loinard07}.

\subsection{Star Formation}

This is a complex subject which has perplexed astronomers for many
years. It is clear that all the answers don't come from only studying
the stars or only studying the newly formed stars. The approach is a
complex multi-wavelength analysis using all the constraints on a
variety of different scales that star formation operates on.  There
is still much to do in our Galaxy. Star formation typically occurs in
regions deeply embedded in dusty molecular gas clouds 
(e.g., young clusters, BCDs etc)
and one of the
highly desirable observations is to observe these processes in detail in a
variety of environments. It is only by observing different physical
conditions that we can hope to fully understand the different
processes that dominate star formation.
These kind of observations require
a sensitive IR imaging telescope. Herschel is making a start, and JWST 
will add to this but E-ELT should have superior spatial resolution and
sensitivity. 

\subsubsection{Young Star Clusters}

Young clusters are extremely small dense systems, 
with central surface densities up to 1000 objects per square 
arcsecond and they are typically
hidden behind and within large amounts of
dust, producing up to $\sim$ 200 mag of visual extinction. 
At present, with HST/WFC3 (but up to 1.6 $\mu$m),  it is only possible 
to look into a few of the nearby bright, less extincted 
($A_V \sim$ 50-100 mag) and less crowded young clusters, which are  
at a late stage of cluster dynamical evolution. 
In the future, 
with JWST this will be extended to 
near- and mid-IR wavelengths. To study a variety 
of these systems in a range of different physical conditions 
throughout the MW and beyond, deep IR adaptive optics imaging with 
the E-ELT at a quasi-diffraction limited resolution of 10 mas 
in the $K$-band (15 mas and 25 mas in the $L$- and 
$M$-band, respectively) is required.
Follow-up integral field spectroscopy will also be important to study 
dynamical processes associated with the cluster formation, such as 
tight binary formation and gravitational interactions followed 
by stellar ejections. 

  \begin{figure}
  \centering
\resizebox{\hsize}{!}{\includegraphics{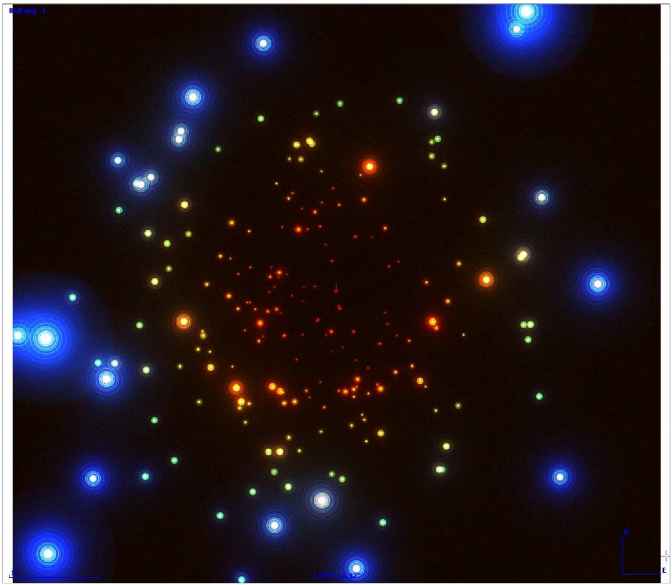}}
     \caption{From A. Calamida, $K,L,M$ composite image of the simulation of a dense massive cluster in the Galactic center as would be observed by E-ELT.}
        \label{cala-fig}
  \end{figure}

Detailed simulations have been carried out (by A. Calamida, 
see Figure~\ref{cala-fig}) to predict
what (and how) we may be able to observe deeply embedded massive stars 
and protostars just formed in dense Galactic protocluster clouds 
(ultra-compact HII regions, hot cores, outflow and maser sources). 
The main restriction on this is spatial resolution 
and sensitivity. 
The imaging simulations confirmed that in order to penetrate an
extinction as high as $A_V$ = 200 mag we will have to observe in the
$L$ and $M$-band.  With the E-ELT and a total exposure time of
$\approx$ 1~hr (+ {\it overheads}), we will be able to observe deeply
embedded ($A_V$ = 200 mag) massive dense clusters at the distance of
the Galactic center (GC), i.e. $\sim$ 7 kpc.
The completeness of the
$L, L-M$ Colour-Magnitude Diagram (CMD) is 25\% but this would be 
sufficient to
determine the stellar number density of the cluster adopting $M$-band
star counts. We will be able to observe massive
dense clusters in the GC in the case of a uniform extinction
distribution of $A_V$ = 150 mag, in the $K,L,M$ bands, with a total
exposure time of $\approx$ 25~hrs (+ {\it overheads}), reaching the
faintest magnitudes ($K \sim$ 29-29.5 mag) with $S/N \sim$ 10.  The
completeness of the three colour CMDs is 100\% and we would be able to
study the stellar number density of the cluster and to probe the
presence of hot and warm circumstellar matter (disks/envelopes) by
means of the detection of infrared excess (in the colours $K-L$ and
$L-M$). If a clumpy extinction is present in the cluster, the
completeness of the CMDs will decrease according to the scale length
of the extinction clumps. If the assumed scale is 0.25" at the GC
distance, we will be able to recover 34\% of stars in the $K$-band,
64\% in the $L$-band and 97\% in the $M$-band with $S/N >$ 5. In the
case of a Gaussian extinction distribution, the recovered
fractions will be $>$ 90\% for the three IR bands.  The presence of
mass segregation at the level of 50\% of stars with masses $>$ 20-30
$M_{\odot}$ being in the inner cluster core does not affect the result
of the simulations.  We will then be able to detect the presence of
mass segregation in a dense massive cluster at the distance of the GC,
if the extinction is less than $A_V$ = 200 mag. 

Finally we mention that ALMA will have a similar spatial
resolution (10 mas)
at sub-mm wavelengths as the E-ELT in the near- and thermal
infrared. Hence, while E-ELT will be particularly good at detecting
$OB$ stars at the end of their main accretion phase, ALMA should see
$OB$ protostars, i.e. massive stellar objects in their early main mass
assembly phase.  Together E-ELT and ALMA are a powerful combination to
reveal one of the most hidden but most important secrets of stellar
astrophysics: the origin of massive stars. 
SKA will be capable of
imaging the thermal, ionised gas at sub-arcsec resolution. The thermal 
radio emission allows an extinction-free measurement of the total input
of ionising photons in the star formation region.

\subsection{Variable Stars}

\subsubsection{Population Indicators}

Despite ever improving star formation histories of Local Group dwarf
galaxies from CMD analysis, many uncertainties remain in the identification and
characterization of the oldest stellar populations. This is because
the old stellar populations can be hard to interpret, or even detect,
due to their inherent faintness and scarcity, and often a strong
overlying young population makes the crowding due to much brighter
stars difficult to overcome. Despite crowding and faintness variable
stars can be relatively easy 
to pick out in the crowded images and their light curves accurately
determined. The easy detection and identification of certain classes of
variable stars help our understanding of the SFH at
ages which are difficult to retrieve from CMD 
analysis alone.

As an example, RR Lyrae variable stars are all $> 10-12$~Gyr 
old and amongst the easiest (and brightest) ancient stars to
unambiguously identify and study. They are in fact low mass stars (M
$\le$ 0.8 M$_{\odot}$) belonging to the Horizontal Branch. They
are about 3 mag brighter than their old MSTO counterparts, which 
makes them easy to unambiguously select 
for further study.
The Leo~A dwarf Irregular galaxy has been the subject of a deep
CMD study with HST/ACS \citep{cole07},
providing an accurate SFH using 
main sequence turnoffs. There appears to have been only a
small and uncertain amount of star formation in Leo A at the earliest
times. The star formation rate at the oldest times 
determined from CMD analysis alone is clearly not
very reliable. However, there is no
doubt that there is an ancient population, 
due to the presence of RR Lyrae variable stars
\citep{Dolphin02}.  
RR Lyrae have also been discovered
in the nearby elliptical galaxy M32 \citep[see][]{fiorentino10a}.
This is an extreme case, where due to the prohibitive crowding
(Monachesi et al., submitted) ancient main sequence turnoffs could not
be detected. Here again, variables 
represent the only way to constrain the 
presence of a stellar population $>10$ Gyrs old in this galaxy. 
Today the photometric detection
limit for the HB is 2.5 Mpc with HST. 
This limit will be extended to 6.3 Mpc with
E-ELT and will allow us to extend the variable star detections to the
whole Local Universe \citep{kara04}.

  \begin{figure}
  \centering
\resizebox{\hsize}{!}{\includegraphics{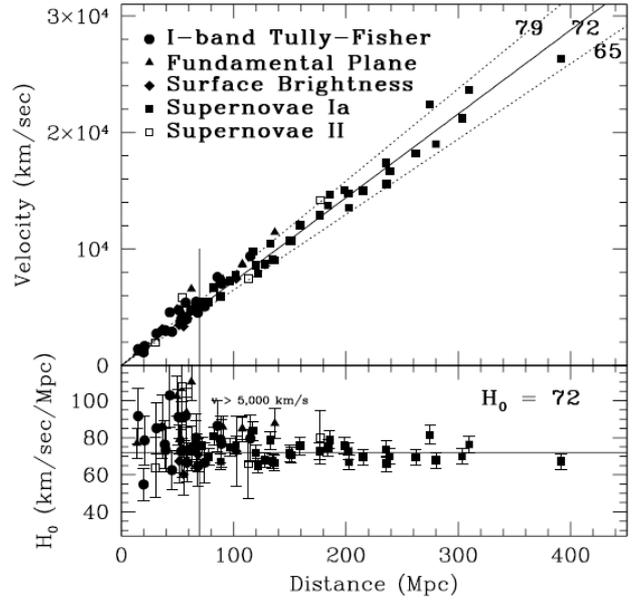}}
     \caption{Hubble diagram of distance vs. velocity for secondary distance indicators calibrated by Cepheids.  A slope of H$_0$ is 72 shown, flanked by $\pm$10\% lines. Bottom: Value of H$_0$ as a function of distance \citep{freedman01}.
}
        \label{Freed-fig}
  \end{figure}

\subsubsection{The distance scale}

RR Lyrae are very robust distance indicators 
due to their very well known 
K-band period-luminosity relation.
However the most common primary distance indicators are the Classical
Cepheids. Observed in young stellar system these are youngest and most
luminous (M$_I$ up to -5 mag) variable stars and they play a key role
as primary distance indicators in the cosmic distance ladder. Due to
the existence of a Period-Luminosity relation, they allow the
calibration of the SNIa till distances out to 
22 Mpc \citep[e.g.,][]{freedman01}. 
The absolute calibration of the SNeIa luminosity peak, and thus the
H$_0$ measurement, is currently  ``anchored'' to  the Cepheid-based 
distances to 10 nearby galaxies, and for this reason Classical   
Cepheids are the cornerstones for the absolute calibration of the
extragalactic distance scale. 
However, the ``universality'' of the Cepheid
Period-Luminosity relation, and the possibility 
that the slope and zero point are sensitive to chemical composition,
has been highly debated for almost two decades
\citep[e.g.,][]{romaniello09}.
Most of the observational and theoretical efforts 
have concentrated on
galaxies with a metal content ranging from the Small
Magellanic Cloud (SMC) value (Z $\sim$ 0.004) to the most metal-rich
galaxies known (Z $\sim$ 0.04), and very little is known for more
metal-poor systems. The effect 
is still an open issue, 
and leads to 
an overall uncertainty on H$_0$ of up to $\sim$20 \%.

In this framework, a very promising but still largely unexplored class
of standard candles is represented by the ultra long period Cepheids
(ULPs; periods larger than 80 days), recently identified in nearby
star-forming galaxies (LMC, SMC, NGC 6822, NGC 55, NGC 300). First,
because the possible flattening of the PL relation at such long
periods makes the ULP Cepheids better standard 
candles than shorter period Classical Cepheids.  Second, and more
importantly, because ULPs are much brighter (by 2-4 magnitudes) than
typical, shorter period, Classical Cepheids (M$_I \sim -5$ mag)  used
so far to set the extragalactic distance ladder. 
{\it Gaia} will provide trigonometric parallaxes at $\mu$arcsec
accuracy, hence precise direct distances, of the LMC and SMC ULP
Cepheids. 
A preliminary application to currently known ULP Cepheids of the
Wesenheit relation commonly used to derive Cepheid distances suggests
a possible metallicity dependence, accounting for about 0.5 mag (over
1.5 dex metallicity spread) of the scatter observed in the ULP's PL
relation. However, these results are not very secure mostly because
only 20 ULP Cepheids are known so 
far. However, most of these variables are in galaxies with metallicities 
at or 
above the SMC (Z $\sim$ 0.004). The only exception is represented by
the two ULPs recently discovered in the most extremely
metal-poor (Z$=$0.0004, 1/50Z$_{\odot}$) Blue Compact Dwarf (BCD)
galaxy I Zw 18 \citep{fiorentino10b}. 
These are the two lowest metallicity ULPs known so far,
and also the most metal-poor Classical Cepheid.

ULP Cepheids observed with HST can be used as ``stellar standard
candles'' to measure {\it directly} distances up to 100 Mpc.  This is
several times the current observational limit \citep[22 Mpc:
][]{freedman01} Cepheid and in the unperturbed ``Hubble Flow''
domain. 
The ULP's potential will be extensively 
enhanced by E-ELT. On the basis of E-ELT simulations,
under the reasonable hypothesis that variable stars
can be measured as faint as 1 mag above the detection limit, assuming a 1 hour
exposure time, it has been shown that ULPs will be observable out to 
a distance of 320 Mpc. This will make it  
the first primary distance indicator capable of directly measuring 
H$_0$.
The 
E-ELT will thus be able to play an
unprecedented role in the definition of the Cosmic Distance Ladder.
Distances affect all aspects of the interpretation of observations of galaxies, from the true mass of young stars, to the correct interpretation of the 
gas and stellar properties.

\section{Conclusions}

HST and
VLT led to astonishing breakthroughs over the last few decades, by 
going to higher spatial
resolution and increasing sensitivity limits. ELT and SKA will undoubtedly
do the
same.

By combining SKA and ELT observations of a range of galaxy 
types in the nearby universe ($<$20Mpc) we can both focus on 
individual objects and look at the global properties – for a 
range of star forming conditions, allowing variation in: 
metallicity, density, environment, dark matter content, rotation, 
large scale kinematic structures and really start to understand 
HOW and WHY stars form on a range of scales,
which is an important aspect to 
allow accurate interpretation of both CMDs from resolved 
stellar populations and faint fuzzy blobs at high redshift. 
We also will have the chance to make exceptionally detailed studies of the
magnetic fields in a different range of galaxy types.

By studying galaxy with only gas observations or only stellar observations will always give an incomplete picture of evolutionary processes

\begin{acknowledgements}
We gratefully acknowledge significant input from Tom Oosterloo and Elias Brinks in defining the nearby universe SKA potential; Hans Zinnecker and Mario Nonino in the young stellar clusters ELT science case.
\end{acknowledgements}

\end{document}